\documentclass[prl,twocolumn,superscriptaddress]{revtex4}
\usepackage{epsfig}
\usepackage{times}
\usepackage{amsmath}
\usepackage{amsfonts}
\usepackage{amssymb}
\usepackage[usenames]{color}
\usepackage{graphicx}

\newcommand{\beq}{\begin{equation}}
\newcommand{\eeq}{\end{equation}}
\newcommand{\beqa}{\begin{eqnarray}}
\newcommand{\eeqa}{\end{eqnarray}}

\begin{document}
\title{Quantum Zeno effect in an unstable system with NMR}
\author{Yuichiro Matsuzaki}
   \affiliation{
    NTT Basic Research Laboratories, NTT Corporation, 
    Kanagawa, 243-0198, Japan
   }
 \author{Hirotaka Tanaka}
    \affiliation{
    NTT Basic Research Laboratories, NTT Corporation, 
    Kanagawa, 243-0198, Japan
    }

\begin{abstract}
We theoretically propose a scheme for verification of quantum
 Zeno effect (QZE) to suppress a decay process with Nuclear Magnetic Resonance (NMR). Nuclear spins are affected by
 low frequency noise, and so one can naturally observe non-exponential
 decay behavior, which is prerequisite in observing QZE.
 We also describe that a key component for QZE, namely measurements on
 the nuclear spin,
 can be realized with NMR in the current technology
 by using non-selective
 architecture of a measurement process.
\end{abstract}

\maketitle

Quantum Zeno effect (QZE) is one of the interesting phenomena of quantum
mechanics.
By performing
frequent measurements to confirm whether it decays
or not,
one can slow or even freeze the decay dynamics of unstable system. 
This counterintuitive phenomena has received many attention since mathematical
foundation of QZE was established by Misra and Sudarshan \cite{BECG01a}.
Besides such academic interest, QZE has many potential applications for
quantum information processing if realized.
QZE is expected to control several decoherence
\cite{facchi2002quantum}.
 Furthermore, there are theoretical proposals to prepare pure states
 \cite{nakazato2003purificationetal} and generate an entanglement
 \cite{nakazato2004preparationetal, fujii2010anti} via QZE. Also, it would be possible to increase a
 success probability for logic gates by using QZE
 \cite{franson2004quantumetal, huang2008interaction}.
 \textcolor{black}{
It is worth mentioning that, although pulse techniques such as a spin
 echo approach
\cite{hahn1950spin} are commonly used to suppress decoherence, they are
different from QZE. Such pulse techniques rely on implementing unitary
evolutions while QZE uses non-unitary evolution, namely projective
measurements. Therefore, verification of QZE still has a fundamental importance to
understand the quantum mechanics and to
 realize quantum information processing .
}

In order to observe quantum Zeno effect, it is
essential to perform such measurements when the unstable system shows a
non-exponential decay
such as a
quadratic decay \cite{NakazatoNamikiPascazio01aetal}. Projective measurements can easily
affect the dynamics of an unstable system with non-exponential decay,
while quantum Zeno effect does not occur for an
exponential decay system via projective measurements \cite{NakazatoNamikiPascazio01aetal}.
Although such a quadratic decay is a consequence of
general features of the Schrodinger equation, it is difficult to observe
such a decay in the experiment. The time region to show the
quadratic decay is usually much shorter than the typical time resolution
of an experimental apparatus, and after showing a quadratic decay, the
unstable system typically shows an exponential decay
\cite{NakazatoNamikiPascazio01aetal}.

Due to the difficulty of observing non-exponential decay, there was
only one experimental demonstration of QZE for an unstable system \cite{FischerGutierrezRaizen01aetal} where the number of
trapped cold sodium atoms are frequently measured to suppress an
escaping rate of the atoms from the trap. It is worth mentioning that,
other than this experiment, not decay process but unitary evolution is suppressed
in previous demonstration of QZE \cite{itano1990quantumetal,StreedMunBoydGretchenCampbellMedleyKetterlePritchard01aetal,xiao2006nmr,
alvarez2010zenoetal}.
Although it is experimentally easier to change the behavior of
unitary evolution by measurements, such approach is different from the
original proposals to suppress irreversible decay process \cite{BECG01a,
FacchiNakazatoPascazio01aetal, Cook01a}.
\textcolor{black}{Moreover, QZE in a
two-level system to suppress decoherence has not been experimentally demonstrated
yet. For the realization of quantum information processing, the use of 
well-defined two-level systems, namely qubits, is essential. So the demonstration of QZE for decoherence by using
such two-level systems would be crucial for the application of QZE to
the quantum information processing \cite{facchi2002quantum,
 nakazato2003purificationetal,nakazato2004preparationetal,franson2004quantumetal,
 huang2008interaction}.
}

Recently, it has been shown that a system affected by low frequency noise is
suitable to observe QZE for an unstable system
\cite{matsuzaki2010quantumzenonttetal}. Due to a long correlation time
of low frequency noise, the decay process naturally shows non-exponential behavior.
 The quadratic decay has been observed experimentally
\cite{KakuyanagiMenoSaitoNakanoSembaTakayanagiDeppeShnirman01aetal,
YoshiharaHarrabiNiskanenNakamura01aetal} in a superconducting qubit affected
by $1/f$ noise.

As suggested in \cite{matsuzaki2010quantumzenonttetal}, a
superconducting qubit affected by $1/f$ noise is one of candidates to observe QZE for an
unstable system. However, 
 the challenge is that a coherence time of the
 superconducting qubit is relatively short such as a few micro seconds
 even at the optimal points \cite{KakuyanagiMenoSaitoNakanoSembaTakayanagiDeppeShnirman01aetal,
YoshiharaHarrabiNiskanenNakamura01aetal}. Also, due to the short coherence time, a high fidelity
 single qubit
 rotation is difficult to be achieved in the current technology. These imperfections result in a significant degradation
 of a signal of QZE \cite{matsuzaki2010quantumzenonttetal}. Therefore, it is preferable to use a system with a longer
 coherence time and a better controllability.

In this paper, to verify QZE of decay process, we suggest a
way to use
Nuclear Magnetic Resonance (NMR).
Nuclear spins have a long coherence time such as a few milli seconds \textcolor{black}{or more}
and
so they are one of the candidates to provide qubits for quantum information
processing. Also, an excellent
control of nuclear spins can be achieved by using a sequence of radio
frequency pulses where the fidelity of the rotation can be more than
$99\%$
in the current technology \cite{ernst1987principles}.
These properties make NMR a strong candidate to demonstrate quantum
information processing in a small number of qubits
\cite{gershenfeld1997bulk}.
Moreover, NMR is affected by a low frequency noise and therefore
naturally shows a non-exponential decay behavior, which is prerequisite in
observing QZE.

To verify QZE with NMR, one of the key components is a
realization of
projective measurements on a single qubit.
In spite of many effort, a projective measurement on a single nuclear
spin with NMR has not been constructed yet. 
Recently, a scheme called spin amplification has been suggested where
the amplitude of a single spin is magnified by using an ensemble of
other spins \cite{divincenzo2000physical, negoro2011scalableetal, close2011rapidetal}, and a gain of 140 in a nuclear system have been
experimentally demonstrated \cite{negoro2011scalableetal}. However, the minimum number of nuclear
spins for detection is the order of $10^6$ and
so a projective measurement of a single nuclear spin is not feasible in the
current technology. In stead of projective measurements, we
here suggest to use non-selective measurements with NMR for the
verification of QZE, as is utilized to stop Rabi oscillations in NMR in
a previous research \cite{xiao2006nmr}. If a pure state is coupled with the other classical
system, non-diagonal terms of the density matrix
disappear due to a decoherence, and this process can be interpreted as a
measurement process without postselection \cite{vNbook32springer,koshino2005quantumshimizu, harris1982two}.
This kind of non-selective architecture of measurement process makes our
proposal feasible even in the current technology.

We summarize the QZE. A fidelity $F=\langle \psi _{\text{ini}}|\rho |\psi _{\text{ini}}\rangle $is utilized to measure the
distance between the decohered state $\rho $ and an initial pure state
$|\psi _{\text{ini}}\rangle $, and we suppress the degradation of the fidelity via frequent measurements.
We performs projective measurements $|\psi
_{\text{ini}}\rangle \langle \psi _{\text{ini}}|$ on the
unstable system with an interval of $\tau =\frac{t}{N}$ where $t$ and $N$ denote a total
time and the number of measurements, respectively. If the fidelity shows a non-exponential decay
without measurements such as
$F\simeq 1-\lambda ^nt^n $ $(1<n\leq 2)$, one can project the state
into $|\psi _{\text{ini}}\rangle $ with a probability $P(N)\simeq (1-\lambda ^n\tau
^n)^N\simeq 1-\frac{\lambda ^nt ^n}{N^{n-1}}$ by performing
$N$ measurements.
As a result, by increasing the number of the
measurements, one can project the state into $|\psi
_{\text{ini}}\rangle $ with almost unity success probability, which is called QZE.
On the other hand, if the dynamics shows an exponential decay, we have
$F\simeq 1-\lambda t $ for the initial stage of the decay. So we
obtain $P(N)\simeq (1-\lambda \tau
)^N\simeq 1-\lambda t$. Since this success probability has no dependency of the number of
measurements, one cannot increase the success probability and therefore
cannot observe QZE for an exponential decay system. 

Let us study a general decoherence dynamics of quantum system.
\textcolor{black}{Although
a decay behavior of unstable system has been studied
\cite{abragam1961principles},
we introduce a simpler solvable model to consider the effect of dephasing induced by classical
noise.}
Typically, the relaxation time is much longer than the dephasing time
for ost of the systems.
For example, 
a typical dephasing
 time in NMR is an order of milli seconds while a relaxation time can be
  more than one hour \textcolor{black}{in some systems} and so the effect of the
 relaxation is negligible.
 \textcolor{black}{ The total Hamiltonian is $H=H_0+H_I$ where $H_0$ and
 $H_I$ denotes the system Hamiltonian and the interaction Hamiltonian
 respectively. We have
  $H_0=\frac{1}{2}\epsilon \hat{\sigma }_z=\frac{1}{2}\epsilon(|1\rangle
 \langle 1|-|-1\rangle \langle -1|)$ where $\epsilon $ denotes an
 energy splitting of the spin.
 The interaction Hamiltonian between the system and the environment is
$H_I=\frac{1}{2}\hat{\eta }(t)\hat{\sigma }_z$ where $\hat{\eta }(t)$
denotes random Gaussian classical noise. We make a unitary
transformation $\tilde{U}=e^{\frac{i}{2}\epsilon \hat{\sigma }_z}$, and
we use the rotating frame at this frequency.
}
Throughout this paper, we consider only
such a system 
in the rotating frame.
Here, we assume that the classical noise $\eta (t)$ is unbiased as
$\overline{\eta (t)}=0$ where the overline denotes the average over the ensemble.
By solving the Schrodinger equation, we obtain
\begin{eqnarray}
\rho (t)-\rho _0=\sum_{n=1}^{\infty
  }(-i)^n\int_{0}^{t}\int_{0}^{t_1}\cdots
  \int_{0}^{t_{n-1}}dt_1dt_2\cdots dt_n\nonumber \\
 \overline{\eta (t_1)\eta (t_2)\cdots \eta (t_n)}[\hat{\sigma }_z,
  [\hat{\sigma }_z, \cdots , [\hat{\sigma }_z,\rho ]\cdots ]]\ \ \ \ \ \
  \ \nonumber 
\end{eqnarray}
where \textcolor{black}{$\rho _0=|\psi _{\text{ini}}\rangle \langle \psi _{\text{ini}}|$ }denotes
a pure initial state.
Since the classical noise $\eta (t)$ is Gaussian, one can decompose 
$\overline{\eta (t_1)\eta (t_2)\cdots \eta (t_n)}$ into a product of two
point correlation functions \cite{Meeron01a}, and so we obtain
\begin{eqnarray}
 \rho (t)&=&\sum_{s,s'=\pm 1}e^{-\frac{1}{4}\Gamma (t)\cdot (s-s')^2 t}
  |s\rangle \langle s|\rho _0|s'\rangle \langle s'| \label{general-form} \\
 \Gamma (t)&=&\frac{1}{2t}\int_{0}^{t}\int_{0}^{t}\overline{\eta
  (t_1)\eta (t_2)}dt_1dt_2\label{decoherence-rate}
\end{eqnarray}
where $\Gamma (t)$ denotes a time dependent decoherence rate.
So, for an initial state $|+\rangle =\frac{1}{\sqrt{2}}|1\rangle
+\frac{1}{\sqrt{2}}|-1\rangle$, the density matrix and the fidelity for
this free induction decay (FID) are
calculated as $\rho _{\text{FID}}(t)=\frac{1+e^{-\Gamma (t)\cdot t} }{2}|+\rangle
\langle +|+ \frac{1-e^{-\Gamma (t)\cdot t} }{2}|-\rangle \langle -|$
and 
$F_{\text{FID}}(t)=\frac{1}{2}(1+e^{-\Gamma (t)\cdot t})$, respectively. Note that, if this
decoherence rate is constant for time, the dynamics here shows an
exponential decay, and
therefore a time
dependency is necessary in this decoherence rate to observe QZE.

\textcolor{black}{
Typically, nuclear spins are affected by $1/f^{\alpha }$ noise where
$0<\alpha <2$, and one of the ways to treat $1/f^{\alpha }$ noise is to
sum up power spectrum density of several Lorentz distributions 
\cite{weissman19881,paladino2002decoherenceetal, galperin2006nonetal,PMVOJK01aetal}.
The spectrum density of a Lorentz distribution is characterized as
 $S_{\text{L}}=\int_{-\infty}^{\infty }\overline{\xi _{\gamma
 }(t)\xi_{\gamma }
   (0)}e^{-i2\pi ft}dt\nonumber \\
  =\frac{\Delta ^2 /\gamma }{1+(\pi f/\gamma )^2}$
 where
 the corresponding correlation function is $ \overline{\xi _{\gamma }(t)\xi _{\gamma }(0)}=\Delta ^2e^{-2\gamma
|t|}$ and $\Delta $ denotes the amplitude of the noise
 \cite{weissman19881,paladino2002decoherenceetal, galperin2006nonetal,PMVOJK01aetal}.
By using
a sum of these 
we have
\begin{eqnarray}
\overline{ \eta 
(t_1)\eta (t_2)}=(\sum_{k=1}^{M}\frac{1}{(\gamma _k)^{\alpha
}}\overline{\xi _{\gamma _k}(t_1)\xi _{\gamma _k}(t_2)})/(\sum_{k=1}^{M}\frac{1}{(\gamma _k)^{\alpha }})\label{many-rtn}
\end{eqnarray}
where $\gamma _k =\gamma _1+(k-1)\delta $, and the power spectral can be approximated to $1/f^{\alpha }$ noise for $\gamma _1 < \pi
f < \gamma _M$ \cite{weissman19881,paladino2002decoherenceetal,
galperin2006nonetal,PMVOJK01aetal}.
}

\begin{figure}[h]
       \begin{center}
        \includegraphics[width=8cm]{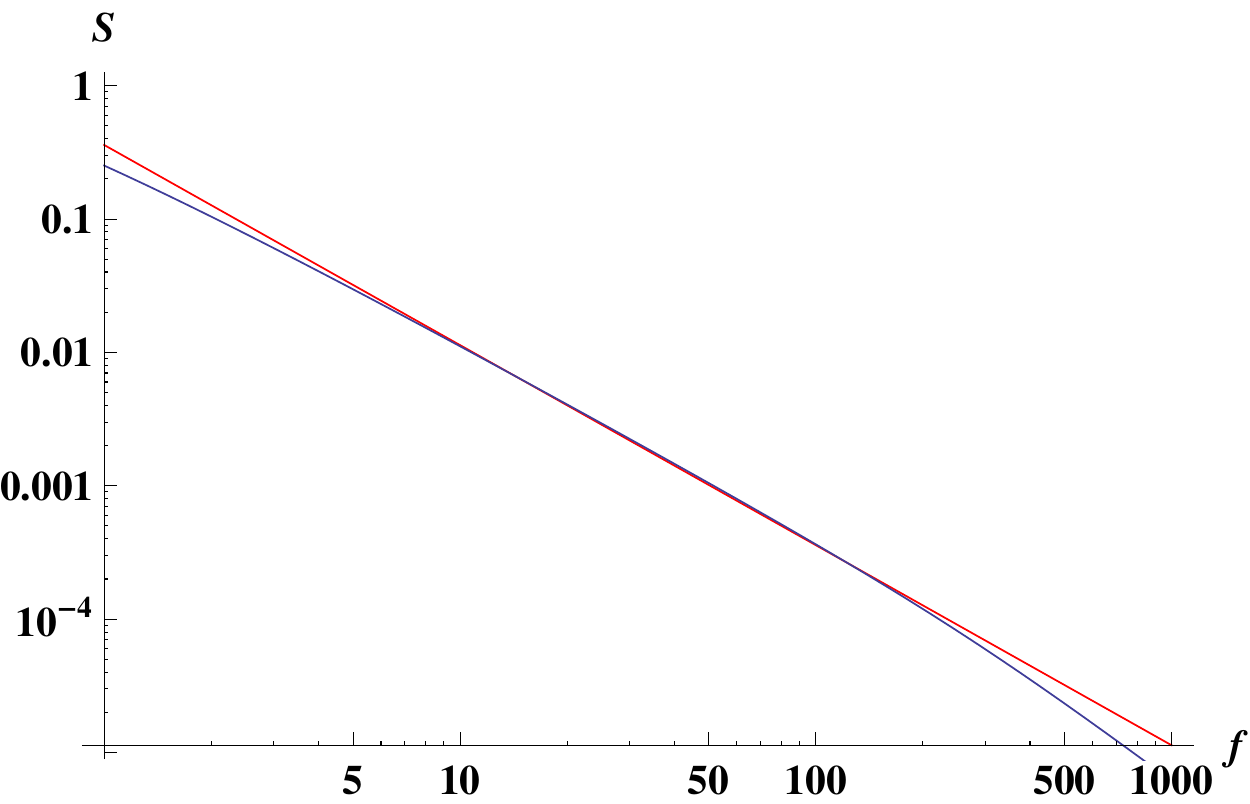} 
        \caption{
        The red line denotes a logarithm of the power spectrum density of $1/f^{1.5}$
        noise while the blue line denotes a logarithm of the power spectrum density of
        \textcolor{black}{a sum of one thousand Lorentz distributions. The
        width of the Lorentz distributions
         are $\gamma _k= k $ (Hz) for
        $(k=1,2,\cdots , 1000)$.} This graph shows that the sum
        of Lorentz distributions can approximately describe $1/f^{1.5 }$ noise
        for $1 \text{ (Hz)}<\pi f< 1000 $ (Hz).
        }\label{noise-spectram}
       \end{center}
       \end{figure}

Since we obtain a general form of decoherence rate under the effect of
classical noise, we can analyze the decoherence dynamics with NMR by
substituting the noise spectrum into (\ref{general-form}) and (\ref{decoherence-rate}).
\textcolor{black}{By using the power spectrum density explained above,}
 we obtain an analytical solution for the time
dependent decoherence rate for $1/f^{\alpha }$ noise as follows.
\begin{eqnarray}
 \Gamma (t)=\Delta ^2(\sum_{k=1}^{M}\frac{1}{(\gamma _k)^{\alpha }
  }\frac{2\gamma _kt -1 +e^{-2\gamma _kt}}{4(\gamma _k)^2t} )/(\sum_{k=1}^{M}\frac{1}{(\gamma _k)^{\alpha }
  })
\end{eqnarray}
\textcolor{black}{Recently, it has been found that} the power spectrum
density in NMR is $1/f^{1.5}$ for
$1 \text{ (Hz)}<\pi f< 1000 $ (Hz), and therefore we use the
sum of Lorentz distributions to model this noise \cite{sasaki}. 
In FIG \ref{noise-spectram}, we plot the power spectrum density of
        a sum of one thousand Lorentz distributions
        and compare this with the power
        spectrum density of $1/f^{1.5}$, which shows that \textcolor{black}{this
        approximation} is quite accurate in this
        frequency range.
                                  \begin{figure}[h]
       \begin{center}
        \includegraphics[width=8cm]{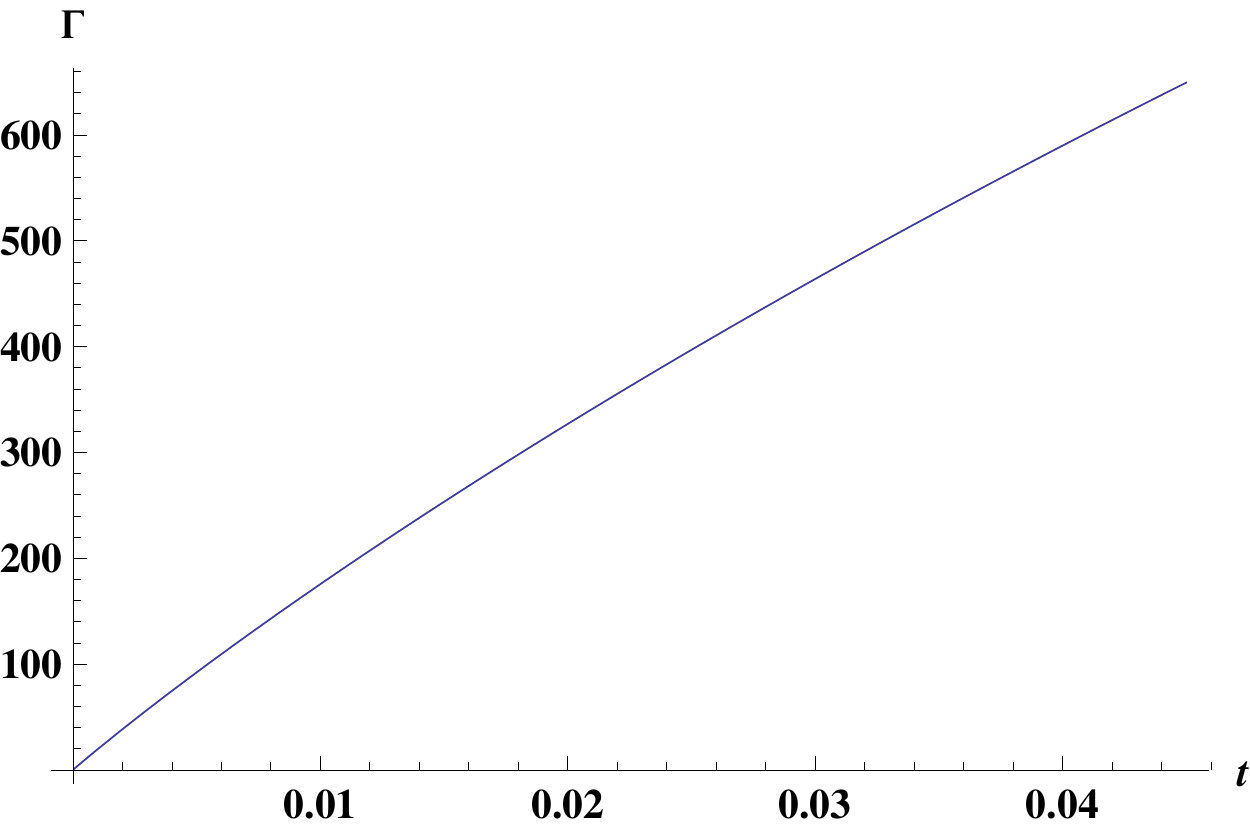} 
        \caption{Decoherence rate to model $1/f^{1.5}$ noise is plotted against time
        where the unit of time is a second.
        The amplitude is $\Delta
        = \frac{1}{T_2}$ where we set $T_2=5$ (ms). Since this
        decoherence rate has a strong time dependency, the dynamics of
        the quantum states naturally
        shows non-exponential decay behavior. \textcolor{black}{It is
        worth mentioning that one can realize the negligible decoherence
        rate $\Gamma \simeq 0$ in the limit of frequent
        measurements, which is called quantum Zeno effect.}
        }\label{decoherence-rate-plot}
       \end{center}
       \end{figure}
Also, we have plotted the decoherence rate in FIG \ref{decoherence-rate-plot}. The plot
        shows a clear time dependency of the decoherence rate, and so
        the decay dynamics here is non-exponential, which is
        prerequisite in observing QZE as explained before.

        We explain how to observe QZE with NMR. Firstly, one prepares an equally
        superposition of $|1\rangle $ and $|-1\rangle$, namely
        $|+\rangle =\frac{1}{\sqrt{2}}(|1\rangle +|-1\rangle )$ (or $|-\rangle =\frac{1}{\sqrt{2}}(|1\rangle -|-1\rangle )$), which
        is an eigenstate of $\hat{\sigma }_x$.
        Secondly, we repeatedly perform measurements about $\hat{\sigma
        }_x$ on a nuclear spin in a time interval $\tau =\frac{t}{N}$.
        Since the dynamics with NMR shows non-exponential decay behavior
        as shown before, the dephasing process will be suppressed via
        the measurements so that the system remains in the initial state.
        However, in this simple scheme, there are two difficulties to be overcome.
        The first difficulty is that, in spite of many effort, a projective measurement \textcolor{black}{on a single nuclear
spin} with NMR has
        not been realized yet. The second difficulty is that it is
        usually much harder to perform a measurement about
        $\hat{\sigma }_x$ than a measurement about $\hat{\sigma }_z$. We
        will discuss how to relax these required conditions.

        Fortunately, to observe QZE, it is
        possible for us to replace the projective measurement with a
        non-selective measurement which is less demanding technology.
        A non-selective measurement denotes a process of a measurement
        without postselection, and this occurs when a principal system
        interacts with other classical systems such as \textcolor{black}{environment}
         but the observer does not obtain
        information about the
        system \cite{vNbook32springer,koshino2005quantumshimizu, harris1982two}.
 When one performs a measurement about $\hat{\sigma }_z$ on a state $\alpha |1\rangle +\beta
 |-1\rangle $ and postselects the measurement result, the
 system is projected into one of the pure states $|-1\rangle $ or
 $|1\rangle $. However, in the non-selective measurements, one discards (or loses ) information of the measurement
 result so that the state should be a classical mixture of
 possible states obtained by the measurements as $\rho =|\alpha |^2
 |1\rangle \langle 1|+|\beta|^2 |-1\rangle \langle -1|$, which is called von
 Neumann mixture \cite{vNbook32springer, koshino2005quantumshimizu}.
 Since it is not necessary to know the measurement result but necessary
 just to interact the principal system with the measurement apparatus, this non-selective
 measurement is much easier to be realized than a projective
 measurement.

\textcolor{black}{ A pulsed magnetic field gradient is one of the ways to perform
  non-selective measurements 
  \cite{ xiao2006nmr}.
  Depending on the position of the spins, different phases are embedded
  which causes a suppression of the Rabi oscillations \cite{xiao2006nmr}
  in NMR, and it should be possible to use such magnetic field gradients to
  suppress decoherence in our scheme.}
 Performing a spin amplification technique
 \cite{divincenzo2000physical, negoro2011scalableetal,
 close2011rapidetal} is also a way for non-selective measurements.
 A pure state of the nuclear spin $\alpha |1\rangle +\beta
  |-1\rangle $ interacts with the other ancillary nuclear spins and evolves into an
 entangled state as $\alpha |1\rangle |11\cdots 1\rangle _{\text{ancilla}} +\beta
 |-1\rangle |-1-1\cdots -1\rangle  _{\text{ancilla}}$ which generates a signal
 much larger than a single nuclear spin.
 Although experimentally demonstrated gain of the signal with NMR via
 the amplification is still 140 \cite{negoro2011scalableetal} which
  is much smaller than the minimum number of nuclear
spins for detection, this amplification is enough to realize
 the non-selective measurement.
  Since the
 entangled state is much more fragile than the state of the single
 spin \cite{dur2004stabilitymacro},
 the entangled state quickly decoheres into a mixed state
 \textcolor{black}{and one obtains the target mixed state as $\rho =|\alpha |^2
 |1\rangle \langle 1|+|\beta|^2 |-1\rangle \langle -1|$. This means that
 the environment effectively observes the nuclear spin and therefore the
 non-selective measurement is induced.}

  \begin{figure}[h]
       \begin{center}
        \includegraphics[width=8.6cm]{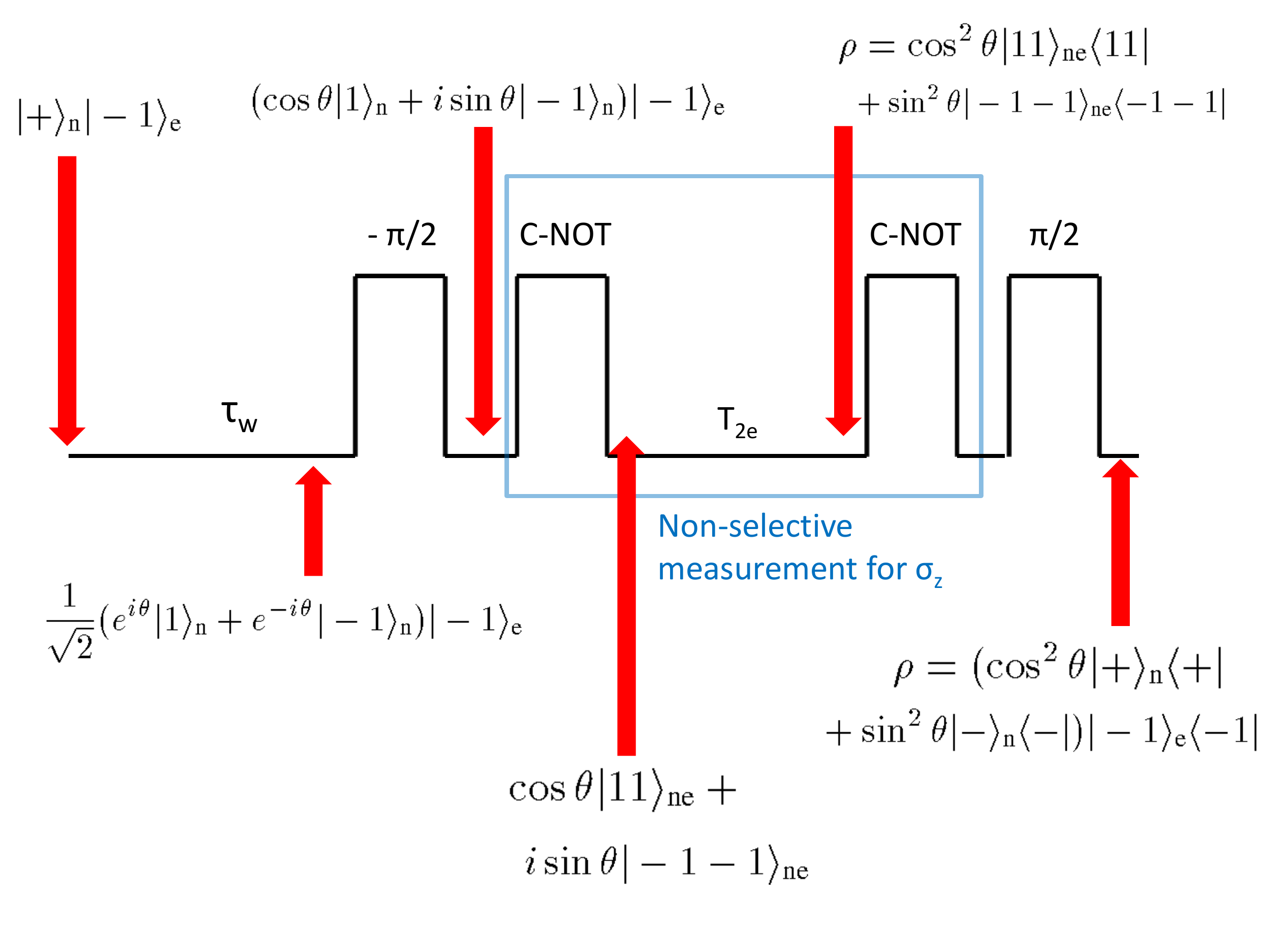} 
        \caption{A pulse sequence to construct a non-selective
        measurement on the nuclear spin via a coupling with an electron
        spin. The nuclear spin state initially has a superposition
        between an excited state and a ground state,
        and this state acquires small unknown phase $\theta $ due to low frequency
        noise.
        A
        controlled-not gate is performed so that
        the nuclear spin is
        entangled with an electron spin, and the composed state
        decoheres as quickly as the electron spin state. 
        In order to separate the electron spin from the nuclear spins,
        another controlled-not gate should be performed.  These
        processes eliminate the non-diagonal term of the density matrix of the
        nuclear spin,
        and therefore can be considered as the non-selective
        measurement to $z$ axis for the nuclear
        spins. If one needs to construct a non-selective measurement
        to $x$ axis, $\pm \frac{\pi }{2}$ pulses on the nuclear should be
        performed before and after the procedure for the $z$ direction measurement.
        }\label{pulse-schematic}
       \end{center}
 \end{figure} 
 Another way to perform a non-selective measurement is to adopt a
 composed system of a nuclear spin and an electron spin \textcolor{black}{as suggested in \cite{xiao2006nmr}}, which can be
 entangled in the current technology
 \cite{simmons2011entanglementetal}. By performing a controlled-not gate
 \textcolor{black}{ on electron nuclear double resonance (ENDOR)}, an initial state $(\alpha |1\rangle _{\text{n}}+\beta
 |-1\rangle _{\text{n}})|-1\rangle _{\text{e}}$ evolves into a Bell state $\alpha |1\rangle _{\text{n}}|1\rangle _{\text{e}}+\beta
 |-1\rangle _{\text{n}}|-1\rangle _{\text{e}}$ \textcolor{black}{where $|\rangle
 _{\text{n}}$  ($|\rangle _{\text{e}}$) denotes the state of the
 nuclear (electron)}, so that the information of
 the nuclear is effectively transfered to the electron.
 Since highly sensitive and local magnetic field sensors have already
been realized and these have potential to perform a projective
 measurement on a single electron spin 
 \cite{rugar2004singleetal,maze2008nanoscaleetal,balasubramanian2008nanoscaleetal}, it might be
 possible in the future to perform a projective measurement on the nuclear spin via
 the entanglement with the electron spin.
 \textcolor{black}{Besides, once
 one generates such an entanglement,
 the nuclear spin is observed by the environment via the dephasing of
 the electron spin. The non-diagonal terms disappears and the state
 becomes $\rho =|\alpha |^2 |1\rangle _{\text{n}}\langle 1|\otimes
 |1\rangle _{\text{e}}\langle 1|+|\beta
 |^2 |-1\rangle _{\text{n}}\langle -1|\otimes |-1\rangle
 _{\text{e}}\langle -1|$.
 In order to separate the electron spin from the nuclear spin, one can
 perform another controlled-not gate, and therefore
  one obtains the target nuclear spin state $\rho =|\alpha |^2
 |1\rangle _{\text{n}}\langle 1|+|\beta|^2 |-1\rangle _{\text{n}}\langle -1|$ as described in
 Fig. \ref{pulse-schematic}.
 }
 This is another feasible implementation of the non-selective measurement on
 the nuclear spin.
 
 \begin{figure}[h]
       \begin{center}
        \includegraphics[width=9cm]{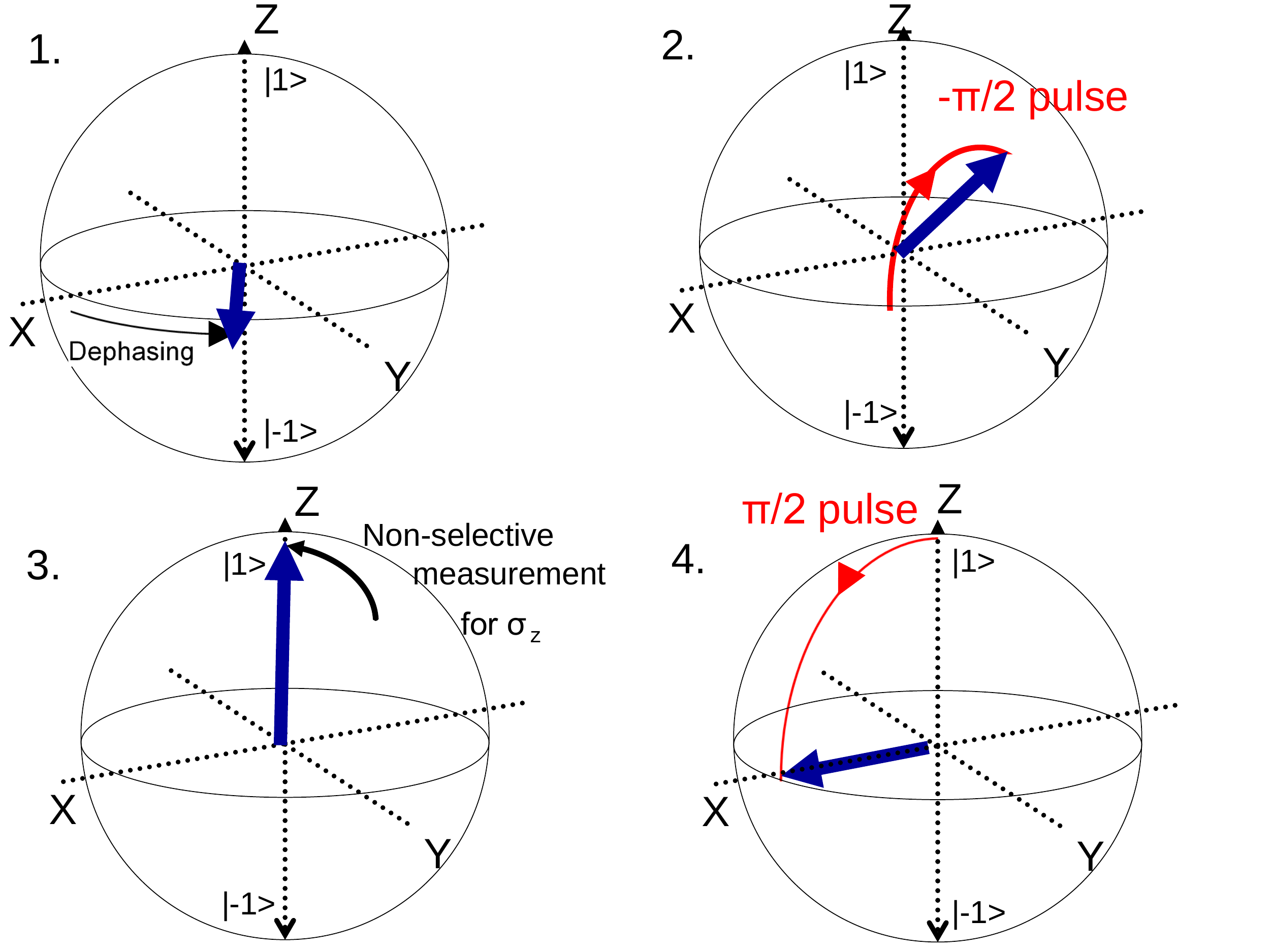} 
        \caption{A schematic of a quantum state in a Bloch sphere to
        verify QZE for a decay process. An initial state $|+\rangle $ is
        freezed by frequent measurements for $\hat{\sigma }_x$.
        In order to construct a
        measurement about $\hat{\sigma }_x$, $\pm \frac{\pi }{2}$ pulses
         are
        performed before and after a measurement about $\hat{\sigma
        }_z$, \textcolor{black}{which are the same
        as 
        illustrated in Fig \ref{pulse-schematic}. }
        Note that a measurement here refers to an interaction with
        another classical system to remove the non-diagonal term of the
        quantum state, which we call a non-selective measurement.        
        }\label{schematic}
       \end{center}
 \end{figure}
 
 Although the implementation of the non-selective measurement mentioned
 above is for $\hat{\sigma }_z$, one can
 easily construct a non-selective measurement about $\hat{\sigma }_x$
 via a sequence of RF pulses.
 Firstly, one performs a $\frac{\pi }{2}$
 rotation $U_y$ around $y$ axis. Secondly, the non-selective measurement
 about $\hat{\sigma }_z$
 is performed. Finally, one performs a rotation
 $U^{\dagger }_y$ around y axis. Note that such a single qubit rotation can
 be performed in tens of micro seconds, much shorter than the
 life time of the nuclear spins, and therefore the effect
 of decay during these process is negligible. 
 As a result, the combination of the three operations results in
$\hat{\mathcal{E}}_{\text{NSMx}}(\rho )=U^{\dagger }_{y}|1 \rangle \langle 1|U_{y} \rho U^{\dagger }_{y}|1\rangle \langle
 1|U_{y}+ U^{\dagger }_{y}|-1\rangle \langle -1|U_{y}\rho U^{\dagger }_{y}|-1\rangle \langle -1|U_{y}=|+ \rangle \langle +|\rho |+\rangle \langle
 +|+ |-\rangle \langle -|\rho |-\rangle \langle -|$, which is equivalent
 to the non-selective measurement of $\hat{\sigma }_x$
 (FIG. \ref{schematic}).
 \textcolor{black}{
 Here, $\hat{\mathcal{E}}_{\text{NSMx}}$ denotes a superoperator for a
 density matrix represented by an operator sum form \cite{NC01b}
}.

                   \begin{figure}[ht]
       \begin{center}
        \includegraphics[width=8cm]{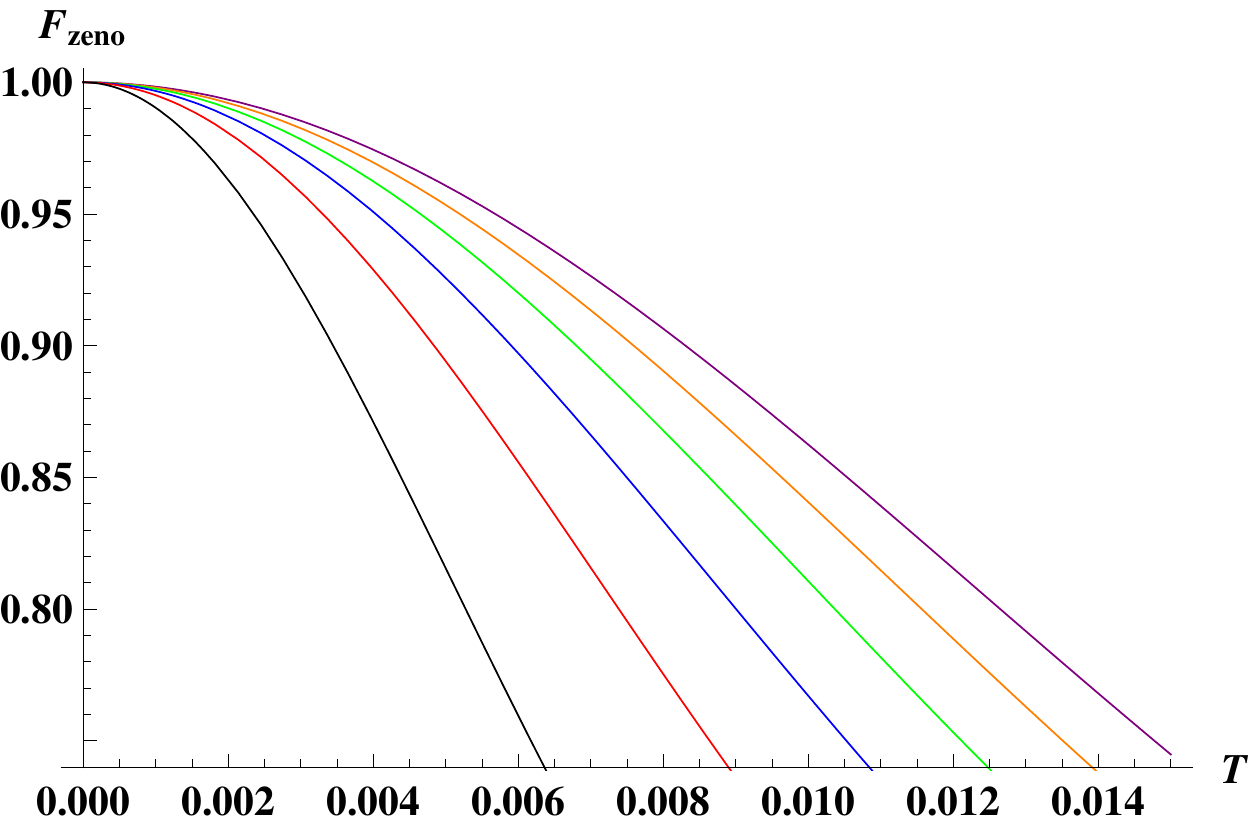} 
        \caption{ Fidelity of quantum states with NMR is against the
        total time
        where the unit of time is a second.
        The first plot from the left denotes a fidelity without
        measurement. The other five plots are fidelities when one performs
        $N=1$, $2$, $3$, $4$, $5$ measurements in a total time $T$,
        respectively.
        The figure clearly shows that the decay is suppressed
        via frequent measurements.
        The noise parameter is the same as in FIG \ref{decoherence-rate-plot}.
        }\label{zeno-fidelity}
       \end{center}
       \end{figure}
                          \begin{figure}[ht]
       \begin{center}
        \includegraphics[width=8cm]{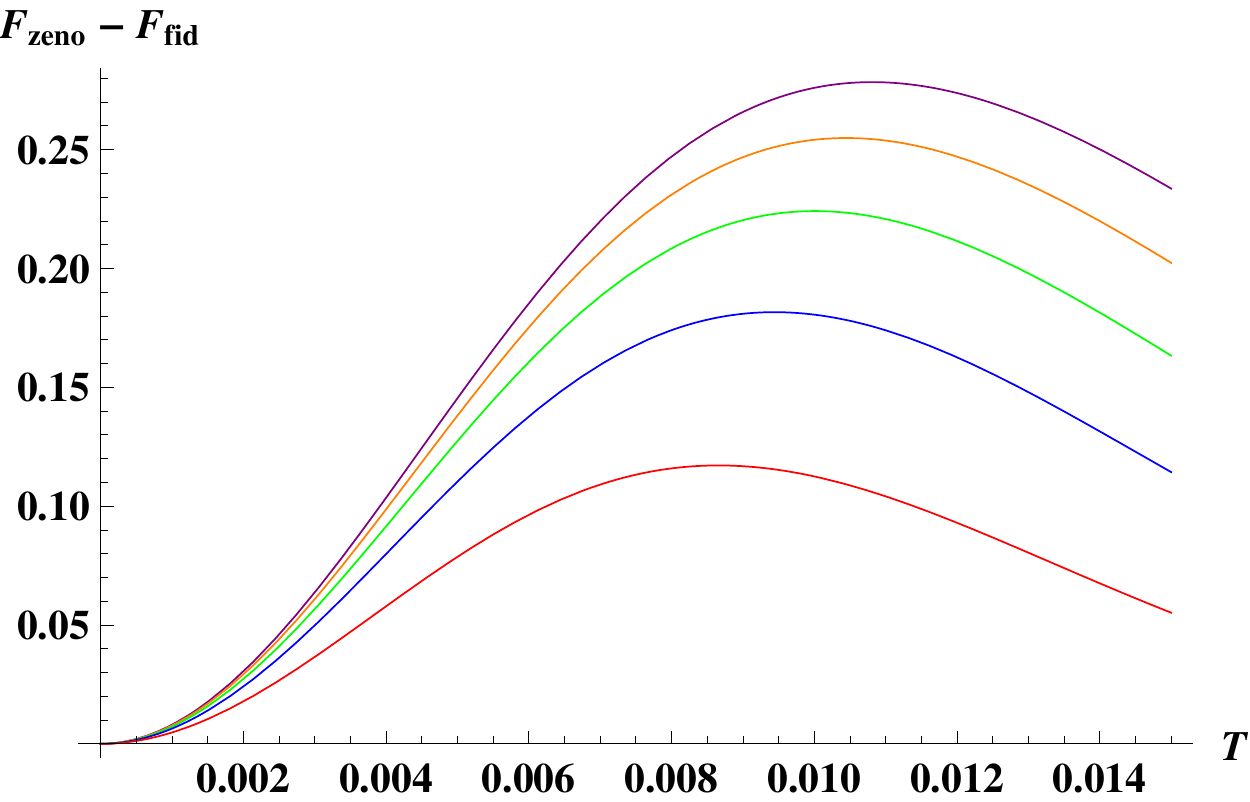} 
        \caption{Difference between  a fidelity with
        measurements and a fidelity 
        without measurement is plotted  against the total time
        where the unit of time is a second.
        The lowest plot denotes a subtraction of a fidelity without
        measurement from a fidelity with a single measurement during the
        total time $T$. The other four plots are subtractions of a fidelity without
        measurement from fidelities when one performs  $N=2,3,4,5$ measurement during the
        total time $T$,
        respectively.
        The parameter of the noise is the same as in FIG \ref{decoherence-rate-plot}.
        }\label{zeno-fidelity-sub}
       \end{center}
       \end{figure}
       We estimate how much QZE can suppress
       the dephasing with NMR compared with the case of the FID.
       We can obtain an analytical form of a density matrix when we perform
       $N$ times of non-selective measurements for a total time $T$ under the effect of low frequency noise described
       by (\ref{many-rtn}) as follows.
       \begin{eqnarray}
        \rho _N&=&\hat{\mathcal{E}}_{\frac{T}{N+1}
         }\hat{\mathcal{E}}_{\text{NSMx}} \cdots
         \hat{\mathcal{E}}_{\text{NSMx}} \hat{\mathcal{E}}_{\frac{T}{N+1}
         }\hat{\mathcal{E}}_{\text{NSMx}} \hat{\mathcal{E}}_{\frac{T}{N+1}
         }(|+\rangle \langle +| ) \nonumber \\
         &=&\frac{1+e^{-\Gamma (\frac{T}{N+1})\cdot T} }{2}|+\rangle
\langle +|+ \frac{1-e^{-\Gamma (\frac{T}{N+1}) \cdot T} }{2}|-\rangle
\langle -|\ \ \ \ \ \ 
       \end{eqnarray}
       Here, $\hat{\mathcal{E}}_{\frac{T}{N+1}}$ denotes a superoperator to
       induce the dephasing of $1/f^{1.5}$ for a period
       $\frac{T}{N+1}$ and $\Gamma (t)$ denotes a time dependent decoherence rate define in
       (\ref{decoherence-rate}). 
       So the fidelity is calculated as
       $F_{\text{QZE}}(T,N)=\langle +|\rho _N|+\rangle =\frac{1+e^{-\Gamma
       (\frac{T}{N+1})\cdot T} }{2}$. We plot the fidelity in FIG
       \ref{zeno-fidelity} and this clearly shows the suppression of the
       dephasing via frequent measurements.
       Also, in FIG
       \ref{zeno-fidelity-sub}, we compare this fidelity
       with the fidelity $F_{\text{FID}}(T)$ for the FID. This shows
       that even performing a single
       measurement increases the fidelity up to $11\%$ and performing $5$
       measurements can increase the fidelity up to $25\%$, compared
       with the fidelity for the free induction decay. These result show
       the feasibility of our proposal even in  the current technology.
       
In conclusion, we suggest a scheme to verify QZE of a decay process with NMR.
 We show that the decay
 behavior of quantum states with NMR is non-exponential, and therefore the decoherence
 dynamics with NMR is
 easily affected by measurements. 
  Moreover, a key component for QZE, namely a measurement on the nuclear spin, can be constructed with existing technology of
 NMR by using non-selective
 architecture. Our proposal is feasible in the current technology.



\begin{thebibliography}{37}
\expandafter\ifx\csname natexlab\endcsname\relax\def\natexlab#1{#1}\fi
\expandafter\ifx\csname bibnamefont\endcsname\relax
  \def\bibnamefont#1{#1}\fi
\expandafter\ifx\csname bibfnamefont\endcsname\relax
  \def\bibfnamefont#1{#1}\fi
\expandafter\ifx\csname citenamefont\endcsname\relax
  \def\citenamefont#1{#1}\fi
\expandafter\ifx\csname url\endcsname\relax
  \def\url#1{\texttt{#1}}\fi
\expandafter\ifx\csname urlprefix\endcsname\relax\def\urlprefix{URL }\fi
\providecommand{\bibinfo}[2]{#2}
\providecommand{\eprint}[2][]{\url{#2}}

\bibitem[{\citenamefont{Misra and Sudarshan}(1977)}]{BECG01a}
\bibinfo{author}{\bibfnamefont{B.}~\bibnamefont{Misra}} \bibnamefont{and}
  \bibinfo{author}{\bibfnamefont{E.~C.~G.} \bibnamefont{Sudarshan}},
  \bibinfo{journal}{J. Math. Phys.} \textbf{\bibinfo{volume}{18}},
  \bibinfo{pages}{756} (\bibinfo{year}{1977}).

\bibitem[{\citenamefont{Facchi and Pascazio}(2002)}]{facchi2002quantum}
\bibinfo{author}{\bibfnamefont{P.}~\bibnamefont{Facchi}} \bibnamefont{and}
  \bibinfo{author}{\bibfnamefont{S.}~\bibnamefont{Pascazio}},
  \bibinfo{journal}{Phys. Rev. Lett.} \textbf{\bibinfo{volume}{89}},
  \bibinfo{pages}{80401} (\bibinfo{year}{2002}).

\bibitem[{\citenamefont{Nakazato and {\it{et
  al}}}(2003)}]{nakazato2003purificationetal}
\bibinfo{author}{\bibfnamefont{H.}~\bibnamefont{Nakazato}} \bibnamefont{and}
  \bibinfo{author}{\bibnamefont{{\it{et al}}}}, \bibinfo{journal}{Phys. Rev.
  Lett.} \textbf{\bibinfo{volume}{90}}, \bibinfo{pages}{60401}
  (\bibinfo{year}{2003}).

\bibitem[{\citenamefont{Nakazato and {\it{et al}
  }}(2004)}]{nakazato2004preparationetal}
\bibinfo{author}{\bibfnamefont{H.}~\bibnamefont{Nakazato}} \bibnamefont{and}
  \bibinfo{author}{\bibnamefont{{\it{et al} }}}, \bibinfo{journal}{Phys. Rev.
  A} \textbf{\bibinfo{volume}{70}}, \bibinfo{pages}{012303}
  (\bibinfo{year}{2004}).

\bibitem[{\citenamefont{Fujii and Yamamoto}(2010)}]{fujii2010anti}
\bibinfo{author}{\bibfnamefont{K.}~\bibnamefont{Fujii}} \bibnamefont{and}
  \bibinfo{author}{\bibfnamefont{K.}~\bibnamefont{Yamamoto}},
  \bibinfo{journal}{Phys. Rev. A} \textbf{\bibinfo{volume}{82}},
  \bibinfo{pages}{042109} (\bibinfo{year}{2010}).

\bibitem[{\citenamefont{Franson and {\it{at
  al}}}(2004)}]{franson2004quantumetal}
\bibinfo{author}{\bibfnamefont{J.}~\bibnamefont{Franson}} \bibnamefont{and}
  \bibinfo{author}{\bibnamefont{{\it{at al}}}}, \bibinfo{journal}{Phys. Rev. A}
  \textbf{\bibinfo{volume}{70}}, \bibinfo{pages}{062302}
  (\bibinfo{year}{2004}).

               \bibitem[{\citenamefont{Hahn}(1950)}]{hahn1950spin}
\bibinfo{author}{\bibfnamefont{E.}~\bibnamefont{Hahn}},
  \bibinfo{journal}{Physical Review} \textbf{\bibinfo{volume}{80}},
  \bibinfo{pages}{580} (\bibinfo{year}{1950}).

\bibitem[{\citenamefont{Huang and Moore}(2008)}]{huang2008interaction}
\bibinfo{author}{\bibfnamefont{Y.}~\bibnamefont{Huang}} \bibnamefont{and}
  \bibinfo{author}{\bibfnamefont{M.}~\bibnamefont{Moore}},
  \bibinfo{journal}{Physical Review A} \textbf{\bibinfo{volume}{77}},
  \bibinfo{pages}{062332} (\bibinfo{year}{2008}).

\bibitem[{\citenamefont{Nakazato and {\it{et
  al}}}(1996)}]{NakazatoNamikiPascazio01aetal}
\bibinfo{author}{\bibfnamefont{H.}~\bibnamefont{Nakazato}} \bibnamefont{and}
  \bibinfo{author}{\bibnamefont{{\it{et al}}}}, \bibinfo{journal}{Int. J. Mod.
  B} \textbf{\bibinfo{volume}{10}}, \bibinfo{pages}{247}
  (\bibinfo{year}{1996}).

\bibitem[{\citenamefont{Fischer and {\it{et
  al}}}(2001)}]{FischerGutierrezRaizen01aetal}
\bibinfo{author}{\bibfnamefont{M.~C.} \bibnamefont{Fischer}} \bibnamefont{and}
  \bibinfo{author}{\bibnamefont{{\it{et al}}}}, \bibinfo{journal}{Phys. Rev.
  Lett.} \textbf{\bibinfo{volume}{87}}, \bibinfo{pages}{040402}
  (\bibinfo{year}{2001}).

\bibitem[{\citenamefont{Itano and {\it{et al}}}(1990)}]{itano1990quantumetal}
\bibinfo{author}{\bibfnamefont{W.}~\bibnamefont{Itano}} \bibnamefont{and}
  \bibinfo{author}{\bibnamefont{{\it{et al}}}}, \bibinfo{journal}{Physical
  Review A} \textbf{\bibinfo{volume}{41}}, \bibinfo{pages}{2295}
  (\bibinfo{year}{1990}).

\bibitem[{\citenamefont{Streed and {\it{et
  al}}}(2006)}]{StreedMunBoydGretchenCampbellMedleyKetterlePritchard01aetal}
\bibinfo{author}{\bibfnamefont{E.~W.} \bibnamefont{Streed}} \bibnamefont{and}
  \bibinfo{author}{\bibnamefont{{\it{et al}}}}, \bibinfo{journal}{Phys. Rev.
  Lett.} \textbf{\bibinfo{volume}{97}}, \bibinfo{pages}{260402}
  (\bibinfo{year}{2006}).

\bibitem[{\citenamefont{Xiao and Jones}(2006)}]{xiao2006nmr}
\bibinfo{author}{\bibfnamefont{L.}~\bibnamefont{Xiao}} \bibnamefont{and}
  \bibinfo{author}{\bibfnamefont{J.}~\bibnamefont{Jones}},
  \bibinfo{journal}{Physics Letters A} \textbf{\bibinfo{volume}{359}},
  \bibinfo{pages}{424} (\bibinfo{year}{2006}).

\bibitem[{\citenamefont{{\'A}lvarez and {\it{et
  al}}}(2010)}]{alvarez2010zenoetal}
\bibinfo{author}{\bibfnamefont{G.}~\bibnamefont{{\'A}lvarez}} \bibnamefont{and}
  \bibinfo{author}{\bibnamefont{{\it{et al}}}}, \bibinfo{journal}{Phys. Rev.
  Lett.} \textbf{\bibinfo{volume}{105}}, \bibinfo{pages}{160401}
  (\bibinfo{year}{2010}).

\bibitem[{\citenamefont{Facchi and {\it{et al}
  }}(2001)}]{FacchiNakazatoPascazio01aetal}
\bibinfo{author}{\bibfnamefont{P.}~\bibnamefont{Facchi}} \bibnamefont{and}
  \bibinfo{author}{\bibnamefont{{\it{et al} }}}, \bibinfo{journal}{Phys. Rev.
  Lett.} \textbf{\bibinfo{volume}{86}}, \bibinfo{pages}{2699}
  (\bibinfo{year}{2001}).

\bibitem[{\citenamefont{Cook}(1988)}]{Cook01a}
\bibinfo{author}{\bibfnamefont{R.~J.} \bibnamefont{Cook}},
  \bibinfo{journal}{Phys. Scr.} \textbf{\bibinfo{volume}{T21}},
  \bibinfo{pages}{49} (\bibinfo{year}{1988}).

\bibitem[{\citenamefont{Matsuzaki and {\it{et
  al}}}(2010)}]{matsuzaki2010quantumzenonttetal}
\bibinfo{author}{\bibfnamefont{Y.}~\bibnamefont{Matsuzaki}} \bibnamefont{and}
  \bibinfo{author}{\bibnamefont{{\it{et al}}}}, \bibinfo{journal}{Phys. Rev. B}
  \textbf{\bibinfo{volume}{82}}, \bibinfo{pages}{180518}
  (\bibinfo{year}{2010}).

\bibitem[{\citenamefont{Kakuyanagi and {\it{et al}
  }}(2007)}]{KakuyanagiMenoSaitoNakanoSembaTakayanagiDeppeShnirman01aetal}
\bibinfo{author}{\bibfnamefont{K.}~\bibnamefont{Kakuyanagi}} \bibnamefont{and}
  \bibinfo{author}{\bibnamefont{{\it{et al} }}}, \bibinfo{journal}{Phys. Rev.
  Lett.} \textbf{\bibinfo{volume}{98}}, \bibinfo{pages}{047004}
  (\bibinfo{year}{2007}).

\bibitem[{\citenamefont{Yoshihara and {\it{et
  al}}}(2006)}]{YoshiharaHarrabiNiskanenNakamura01aetal}
\bibinfo{author}{\bibfnamefont{F.}~\bibnamefont{Yoshihara}} \bibnamefont{and}
  \bibinfo{author}{\bibnamefont{{\it{et al}}}}, \bibinfo{journal}{Phys. Rev.
  Lett.} \textbf{\bibinfo{volume}{97}}, \bibinfo{pages}{167001}
  (\bibinfo{year}{2006}).

\bibitem[{\citenamefont{Ernst et~al.}(1987)\citenamefont{Ernst, Bodenhausen,
  Wokaun et~al.}}]{ernst1987principles}
\bibinfo{author}{\bibfnamefont{R.}~\bibnamefont{Ernst}},
  \bibinfo{author}{\bibfnamefont{G.}~\bibnamefont{Bodenhausen}},
  \bibinfo{author}{\bibfnamefont{A.}~\bibnamefont{Wokaun}},
  \bibnamefont{et~al.}, \emph{\bibinfo{title}{Principles of nuclear magnetic
  resonance in one and two dimensions}}, vol. \bibinfo{volume}{332}
  (\bibinfo{publisher}{Clarendon Press Oxford}, \bibinfo{year}{1987}).

\bibitem[{\citenamefont{Gershenfeld and Chuang}(1997)}]{gershenfeld1997bulk}
\bibinfo{author}{\bibfnamefont{N.}~\bibnamefont{Gershenfeld}} \bibnamefont{and}
  \bibinfo{author}{\bibfnamefont{I.}~\bibnamefont{Chuang}},
  \bibinfo{journal}{Science} \textbf{\bibinfo{volume}{275}},
  \bibinfo{pages}{350} (\bibinfo{year}{1997}).

\bibitem[{\citenamefont{DiVincenzo}(2000)}]{divincenzo2000physical}
\bibinfo{author}{\bibfnamefont{D.}~\bibnamefont{DiVincenzo}},
  \bibinfo{journal}{Scalable quantum computers} pp. \bibinfo{pages}{1--13}
  (\bibinfo{year}{2000}).

\bibitem[{\citenamefont{Negoro and {\it{etal}}}(2011)}]{negoro2011scalableetal}
\bibinfo{author}{\bibfnamefont{M.}~\bibnamefont{Negoro}} \bibnamefont{and}
  \bibinfo{author}{\bibnamefont{{\it{etal}}}}, \bibinfo{journal}{Phys. Rev.
  Lett.} \textbf{\bibinfo{volume}{107}}, \bibinfo{pages}{50503}
  (\bibinfo{year}{2011}).

\bibitem[{\citenamefont{Close and {\it{et al}}}(2011)}]{close2011rapidetal}
\bibinfo{author}{\bibfnamefont{T.}~\bibnamefont{Close}} \bibnamefont{and}
  \bibinfo{author}{\bibnamefont{{\it{et al}}}}, \bibinfo{journal}{Phys. Rev.
  Lett.} \textbf{\bibinfo{volume}{106}}, \bibinfo{pages}{167204}
  (\bibinfo{year}{2011}).

\bibitem[{\citenamefont{von Neumann}(1932)}]{vNbook32springer}
\bibinfo{author}{\bibfnamefont{J.}~\bibnamefont{von Neumann}},
  \emph{\bibinfo{title}{Mathematische Grundlagen der Quantenmechanik}}
  (\bibinfo{publisher}{Springer}, \bibinfo{year}{1932}).

\bibitem[{\citenamefont{Koshino and Shimizu}(2005)}]{koshino2005quantumshimizu}
\bibinfo{author}{\bibfnamefont{K.}~\bibnamefont{Koshino}} \bibnamefont{and}
  \bibinfo{author}{\bibfnamefont{A.}~\bibnamefont{Shimizu}},
  \bibinfo{journal}{Phys. Rep.} \textbf{\bibinfo{volume}{412}},
  \bibinfo{pages}{191} (\bibinfo{year}{2005}).

\bibitem[{\citenamefont{Harris and Stodolsky}(1982)}]{harris1982two}
\bibinfo{author}{\bibfnamefont{R.}~\bibnamefont{Harris}} \bibnamefont{and}
  \bibinfo{author}{\bibfnamefont{L.}~\bibnamefont{Stodolsky}},
  \bibinfo{journal}{Physics Letters B} \textbf{\bibinfo{volume}{116}},
  \bibinfo{pages}{464} (\bibinfo{year}{1982}).

                
\bibitem[{\citenamefont{Abragam and Hebel}(1961)}]{abragam1961principles}
\bibinfo{author}{\bibfnamefont{A.}~\bibnamefont{Abragam}} \bibnamefont{and}
  \bibinfo{author}{\bibfnamefont{L.}~\bibnamefont{Hebel}},
  \bibinfo{journal}{American Journal of Physics} \textbf{\bibinfo{volume}{29}},
  \bibinfo{pages}{860} (\bibinfo{year}{1961}).

\bibitem[{\citenamefont{Meeron}(1957)}]{Meeron01a}
\bibinfo{author}{\bibfnamefont{E.}~\bibnamefont{Meeron}}, \bibinfo{journal}{J.
  Chem. Phys.} \textbf{\bibinfo{volume}{27}}, \bibinfo{pages}{67}
  (\bibinfo{year}{1957}).

\bibitem[{\citenamefont{Weissman et~al.}(1988)}]{weissman19881}
\bibinfo{author}{\bibfnamefont{M.}~\bibnamefont{Weissman}}
  \bibnamefont{et~al.}, \bibinfo{journal}{Reviews of modern physics}
  \textbf{\bibinfo{volume}{60}}, \bibinfo{pages}{537} (\bibinfo{year}{1988}).

\bibitem[{\citenamefont{Paladino and {\it{et
  al}}}(2002)}]{paladino2002decoherenceetal}
\bibinfo{author}{\bibfnamefont{E.}~\bibnamefont{Paladino}} \bibnamefont{and}
  \bibinfo{author}{\bibnamefont{{\it{et al}}}}, \bibinfo{journal}{Phys. Rev.
  Lett.} \textbf{\bibinfo{volume}{88}}, \bibinfo{pages}{228304}
  (\bibinfo{year}{2002}).

\bibitem[{\citenamefont{Galperin and {\it{et al}}}(2006)}]{galperin2006nonetal}
\bibinfo{author}{\bibfnamefont{Y.}~\bibnamefont{Galperin}} \bibnamefont{and}
  \bibinfo{author}{\bibnamefont{{\it{et al}}}}, \bibinfo{journal}{Phys. Rev.
  Lett.} \textbf{\bibinfo{volume}{96}}, \bibinfo{pages}{97009}
  (\bibinfo{year}{2006}).

\bibitem[{\citenamefont{P.Kuopanportti and {\it{et al}}}(2008)}]{PMVOJK01aetal}
\bibinfo{author}{\bibnamefont{P.Kuopanportti}} \bibnamefont{and}
  \bibinfo{author}{\bibnamefont{{\it{et al}}}}, \bibinfo{journal}{Phys. Rev. A}
  \textbf{\bibinfo{volume}{77}}, \bibinfo{pages}{032334}
  (\bibinfo{year}{2008}).

                \bibitem[{\citenamefont{S.Sasaki }(2008)}]{sasaki}
\bibinfo{author}{\bibnamefont{S.Sasaki (private communication)}} \bibnamefont{}
  \bibinfo{author}{\bibnamefont{}} \bibinfo{journal}{}
  \textbf{\bibinfo{volume}{}} \bibinfo{pages}{}
  \bibinfo{year}{}.

\bibitem[{\citenamefont{Dur and Briegel}(2004)}]{dur2004stabilitymacro}
\bibinfo{author}{\bibfnamefont{W.}~\bibnamefont{Dur}} \bibnamefont{and}
  \bibinfo{author}{\bibfnamefont{H.}~\bibnamefont{Briegel}},
  \bibinfo{journal}{Phys. Rev. Lett.} \textbf{\bibinfo{volume}{92}},
  \bibinfo{pages}{180403} (\bibinfo{year}{2004}).

\bibitem[{\citenamefont{Simmons and {\it{et al}
  }}(2011)}]{simmons2011entanglementetal}
\bibinfo{author}{\bibfnamefont{S.}~\bibnamefont{Simmons}} \bibnamefont{and}
  \bibinfo{author}{\bibnamefont{{\it{et al} }}}, \bibinfo{journal}{Nature}
  \textbf{\bibinfo{volume}{470}}, \bibinfo{pages}{69} (\bibinfo{year}{2011}).

\bibitem[{\citenamefont{Rugar and {\it {et al}}}(2004)}]{rugar2004singleetal}
\bibinfo{author}{\bibfnamefont{D.}~\bibnamefont{Rugar}} \bibnamefont{and}
  \bibinfo{author}{\bibnamefont{{\it {et al}}}}, \bibinfo{journal}{Nature}
  \textbf{\bibinfo{volume}{430}}, \bibinfo{pages}{329} (\bibinfo{year}{2004}),
  ISSN \bibinfo{issn}{0028-0836}.

\bibitem[{\citenamefont{Maze and {\it {et al}}}(2008)}]{maze2008nanoscaleetal}
\bibinfo{author}{\bibfnamefont{J.}~\bibnamefont{Maze}} \bibnamefont{and}
  \bibinfo{author}{\bibnamefont{{\it {et al}}}}, \bibinfo{journal}{Nature}
  \textbf{\bibinfo{volume}{455}}, \bibinfo{pages}{644} (\bibinfo{year}{2008}),
  ISSN \bibinfo{issn}{0028-0836}.

\bibitem[{\citenamefont{Balasubramanian and {\it{et
  al}}}(2008)}]{balasubramanian2008nanoscaleetal}
\bibinfo{author}{\bibfnamefont{G.}~\bibnamefont{Balasubramanian}}
  \bibnamefont{and} \bibinfo{author}{\bibnamefont{{\it{et al}}}},
  \bibinfo{journal}{Nature} \textbf{\bibinfo{volume}{455}},
  \bibinfo{pages}{648} (\bibinfo{year}{2008}).

\bibitem[{\citenamefont{Nielsen and Chuang}(2000)}]{NC01b}
\bibinfo{author}{\bibfnamefont{M.~A.} \bibnamefont{Nielsen}} \bibnamefont{and}
  \bibinfo{author}{\bibfnamefont{I.~L.} \bibnamefont{Chuang}},
  \emph{\bibinfo{title}{Quantum Computation and Quantum Information}}
  (\bibinfo{publisher}{Cambridge}, \bibinfo{year}{2000}).

\end{thebibliography}

\end{document}